\documentclass[runningheads]{llncs}
\usepackage{graphicx}
\usepackage[T1]{fontenc}
\usepackage{caption}
\usepackage{subcaption}
\usepackage{graphicx}

\usepackage{booktabs}

\begin{document}
\title{diveXplore at the Video Browser Showdown 2024}
\titlerunning{diveXplore 2024 at VBS}

% \author{Andreas Leibetseder\orcidID{0000-0002-9535-966X}
\author{Klaus Schoeffmann \and 
Sahar Nasirihaghighi} 

\authorrunning{K. Schoeffmann et al.}
\institute{Klagenfurt University, Institute of Information Technology (ITEC), \\ Klagenfurt, Austria
\email{\{klaus.schoeffmann,sahar.nasirihaghighi\}@aau.at}}
\maketitle              % typeset the header of the contribution
\begin{abstract}
According to our experience from VBS2023 and the feedback from the IVR4B special session at CBMI2023, we have largely revised the diveXplore system for VBS2024. It now integrates OpenCLIP trained on the LAION-2B dataset for image/text embeddings that are used for free-text and visual similarity search, a query server that is able to distribute different queries and merge the results, a user interface optimized for fast browsing, as well as an exploration view for large clusters of similar videos (e.g., weddings, paraglider events, snow and ice scenery, etc.). 
\end{abstract}


\section{Introduction}

The Video Browser Showdown (VBS) \cite{lokovc2023interactive,lokoc2019interactive,rossetto2020interactive}, which was started in 2012 as a video browsing competition with a rather small data set (30 videos with an average duration of 77 minutes), has evolved significantly since then and became an extremely challenging international evaluation platform for interactive video retrieval tools. It not only uses several datasets with different content, as shown in Table~\ref{tab:videodatasets}, but VBS also evaluates several types of query in different sessions. 

\begin{table}
    \caption{Video datasets used for the VBS}
    \label{tab:videodatasets}
    \centering
    \begin{tabular}{|l|l|r|r|}
    \hline
         \textbf{Dataset} & \textbf{Content} & \textbf{Files} & \textbf{Hours}  \\
         \hline
         V3C1 + V3C2 \cite{berns2019v3c1,rossetto2021insights} & Diverse content uploaded by users on Vimeo & 17,235 & 2,500.00 \\
         Marine videos \cite{MVK} & Diving videos & 1,371 & 12.25 \\
         Surgical videos & Surgery videos from laparoscopic gynecology & 75 & 104.75\\
         \hline
    \end{tabular}
\end{table}

Known item search (KIS) queries require competitors to find a specific segment in the entire video collection, which could be as small as a 2-second clip. 
KIS queries would be issued as \textit{visual} KIS, where the desired clip is presented to the participants via the shared server and/or the on-site projector, or as a \textit{textual} description of the clip, which is even more challenging, as it leaves room for subjective interpretation and imagination. Visual KIS simulates the typical situation where one knows of a specific video segment and knows that it is somewhere contained in the collection, but does not know where to find it. On the other hand, textual KIS simulates a situation where someone wants to find a specific video segment, but does not know how it looks like (or cannot search himself/herself and only describes it to a video retrieval expert). For KIS queries, it is crucial to be as fast and accurate as possible. The faster the clip is found, the more points a team will get. Wrong submissions will be penalized. 
Ad-hoch search (AVS) queries are another type of search where no specific segment needs to be found, but many examples that fit a specific description (e.g., \textit{clips where people are holding a baloon}). For AVS tasks, it is important to find as many examples as possible, whereas diversity in terms of video files will get rated higher. 
Question answering (QA) queries are the final type of search, where answers to specific questions (e.g., \textit{ What is the name of the bride in the wedding on the beach in Thailand?}) need to be sent as plain text by a team (this type of query is new for VBS2024). 

At the VBS, queries are performed by different users. In a first session, the teams themselves (i.e., the video search \textit{experts}) solve various queries for the different datasets. Then, typically the next day, volunteers from the audience (the \textit{novices}) are recruited to use the systems of the teams and solve queries (usually a little easier ones, such as visual KIS and AVS in the V3C datset). 
Therefore, video search systems at VBS have to be very efficient and effective at the retrieval itself, but also fast, flexible, and easy in their usage to both experts and novices. VBS systems have to be designed very carefully and with the many facets of the VBS in mind. 

The Klagenfurt University diveXplore system has been participating in the VBS for several years already \cite{leibetseder2022divexplore,schoeffmann2023divexplore}. However, for VBS2024 it was significantly redesigend and optimized in several components, so that it should be much more competitive than in previous years, while still being easy and effective to use. The most important changes include (i) the integration of OpenCLIP \cite{openclip2022} trained on the LAION-2B \cite{schuhmann2021laion} dataset for image/text embeddings that are used during free-text and similarity search, (ii) the redesign of the query server that is now able to perform and merge parallel queries (which can be temporal queries, or metadata and embeddings combinations, for example), and (iii) user interface optimizations for quick inspection of the context of search results, quick navigation of results, and clustering of similar videos.

\section{diveXplore 2024}
\label{sec:divexplore}
\label{sec:architecture}

\begin{figure}[htb!]
  \centering
  \includegraphics[width=\linewidth]{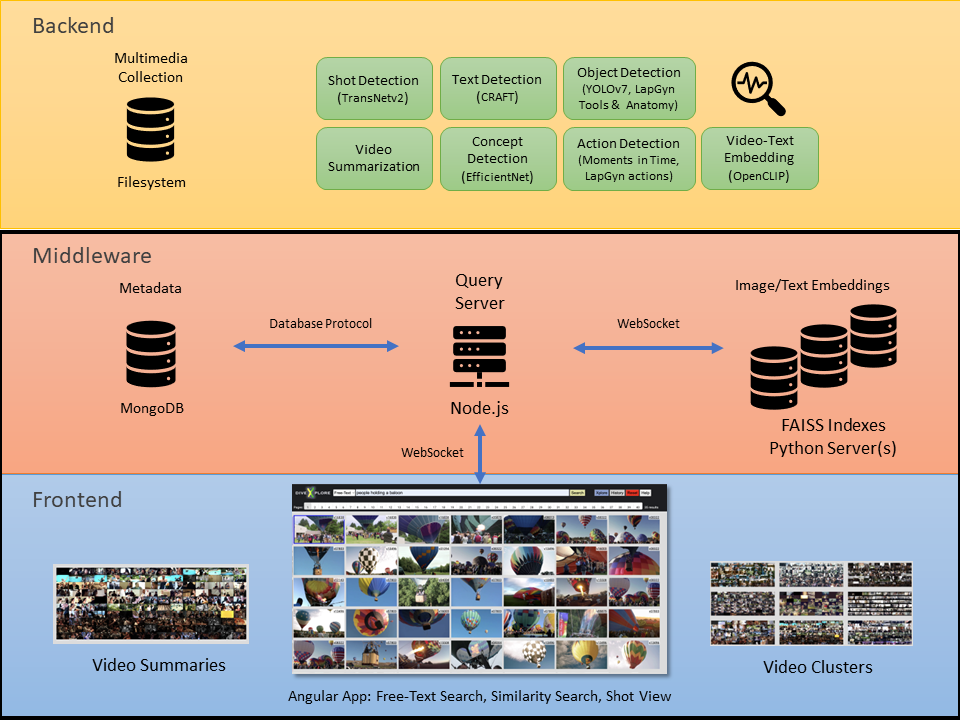}
  \caption{diveXplore 2024 architecture}
  \label{fig:architecture}
\end{figure}

Figure~\ref{fig:architecture} shows the that architecture of diveXplore consists of three major components: the backend, the middleware, and the frontend. Before the system can be used, all videos are analyzed by the backend, which stores image/text embeddings in a FAISS \cite{faiss2019} index, and all other results in the metadata database (with MongoDB\footnote{\url{https://www.mongodb.com}}). The middleware and the frontend are used together and communicate via a WebSocket protocol; while the middleware is responsible for the actual search and retrieval, the frontend is used to present the results and to interact with the user.

\subsection{Video Analysis}

The first step of the analysis is shot segmentation and keyframe extraction, which is done with TransNetv2 \cite{souvcek2020transnet} for V3C and a specific keyframe extraction algorithm for endoscopic videos \cite{schoeffmann2015keyframe} that is based on ORB keypoint tracking. As keyframes, the middle of the shot is used. For videos in the marine dataset, we simply use uniform temporal subsampling, with a 1-second interval. 
The keyframes are then used for further analysis. All keyframes are analyzed with OpenCLIP ViT-H/14 (laion2b) \cite{openclip2022,schuhmann2021laion} and the obtained embeddings are added to a FAISS index \cite{faiss2019}, which is later used by the query server for free text and similarity search. The V3C keyframes are also analyzed for contained objects (COCO), concepts (Places365), text (CRAFT), and actions (Moments in Time), as shown in Table~\ref{tab:analysis} (see also ~\cite{schoeffmann2023divexplore}). The keyframes from the surgery videos are analyzed for surgical tools, anatomical structures, and a few common surgical actions, such as cutting, cauterization, and suturing \cite{nasirihaghighi2023action}. 
Furthermore, all keyframes of a video are used to create video summaries and similarity clusters of videos, as detailed in \cite{schoeffmann2023cbmi}.

\setlength{\aboverulesep}{0pt}
\setlength{\belowrulesep}{0pt}

\begin{table}[htbp!]
    \centering
    \setlength\tabcolsep{2.0pt} % col spacing, default value: 6pt
    \caption{Overview of analysis components used by diveXplore 2024 (V=V3C dataset, M=Marine dataset, S=Surgery dataset)}
    \label{tab:analysis}
    \begin{tabular}{|c|c|c|l|l|c|l|}
        \hline
        \toprule    
        \textbf{V} & \textbf{M} & \textbf{S} & \textbf{Type} & \textbf{Model} & \textbf{Ref.} & \textbf{Description}\\
        \midrule
        x & - & - & Shots & TransNetv2 & \cite{souvcek2020transnet} & Deep model to detect shot boundaries. \\
        - & x & - & Segments & Uniform subsampling & & 1s-segments. \\
        - & - & x & Segments & Endoscopy-Segments & \cite{schoeffmann2015keyframe} & Semantic segments with coherent content.\\
        x & - & - & Concepts & EfficientNet B2 & \cite{tan2019efficientnet,zhou2017places} & Places365 categories. \\
        x & - & - & Objects & YOLOv7 & \cite{wang2022yolov7} & 80 MS COCO categories. \\
        - & - & x & Med.Objects & Mask R-CNN & \cite{kletz2020instrument} & Tools and anatomy in gynecology. \\
        - & - & x & Actions & Bi-LSTM & \cite{nasirihaghighi2023action} & Common actions in gynecology. \\
        x & - & - & Events & Moments in Time & \cite{monfortmoments} & 304 Moments in Time action events. \\
        x & - & x & Texts & CRAFT & \cite{craft2019,craft-rec-2019} & Text region localization and OCR. \\
        x & x & x & Embeddings & OpenCLIP ViT-H/14 & \cite{clip2021} & Embeddings with 1024 dimensions. \\
        x & x & x & Similarity & OpenCLIP ViT-H/14 & \cite{clip2021} & Embeddings with 1024 dimensions. \\
        \bottomrule
        \hline
    \end{tabular}

\end{table}

\subsection{Query Server}

The query server is implemented with Node.js and communicates with the frontend via a websocket connection. Its main purpose is to receive queries from the frontend, analyze these queries, split and forward them, and collect and merge results. This design allows not only for scalability (e.g., to send consecutive queries to alternating index servers), but also to issue several queries in parallel. For example, the query from the frontend could contain free-text that should be forwarded to the FAISS index and an object detected by YOLO that is stored in the metadata database and should be forwarded to MongoDB. Then the query server would wait for results of both queries and merge or filter them, whereas different strategies could be selected for this, e.g., filter the results from FAISS with the video IDs returned by MongoDB, or fuse results from both sources with a particular ranking scheme. Also, the query from the frontend could contain a number of free-text queries that are sent to several FAISS indexes at the same time and the query server would merge the results in a temporal way, i.e., that only results from those videos are finally returned to the frontend that contain matches for all queries in a specific order and in temporal proximity.

\subsection{User Interface}

\begin{figure}[htb!]
  \centering
  \includegraphics[width=\linewidth]{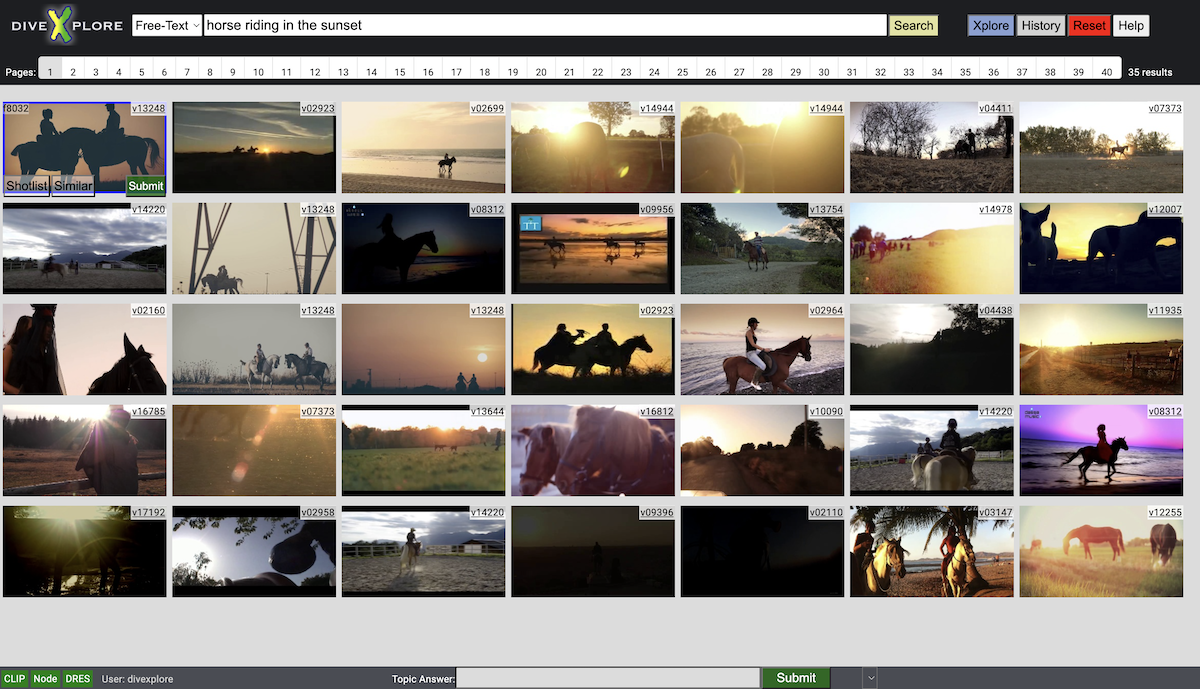}
  \caption{diveXplore User Interface}
  \label{fig:userinterface}
\end{figure}

\begin{figure}
\centering

%\begin{subfigure}{0.49\textwidth}
%  \includegraphics[width=\textwidth]{figures/Textquery.png}
%  \caption{Free-text search interface.}
%  \label{fig:freetextsearch}
%\end{subfigure}
%\hfill

\begin{subfigure}{0.9\textwidth}
  \includegraphics[width=\textwidth]{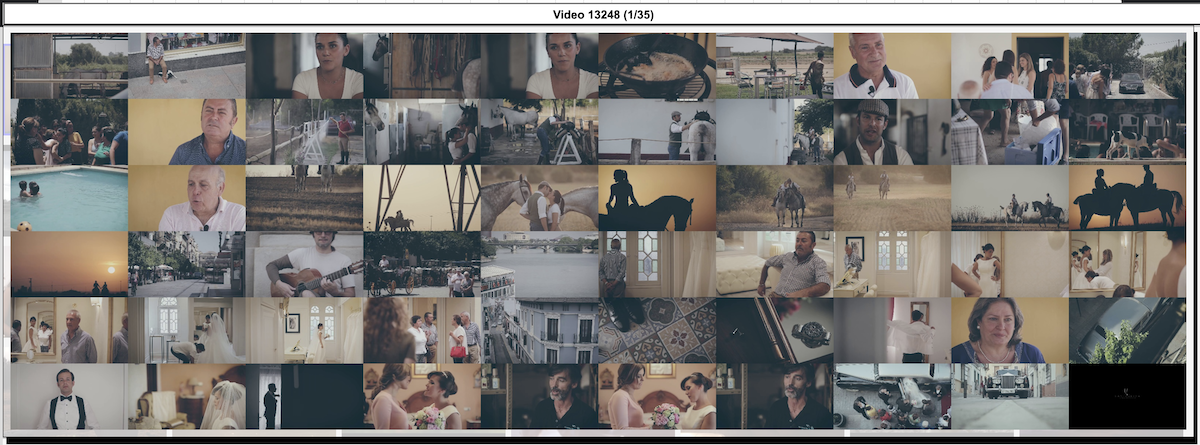}
  \caption{Video summary for quick context inspection.}
  \label{fig:videosummary}
\end{subfigure}
\hfill
\begin{subfigure}{0.9\textwidth}
  \includegraphics[width=\textwidth]{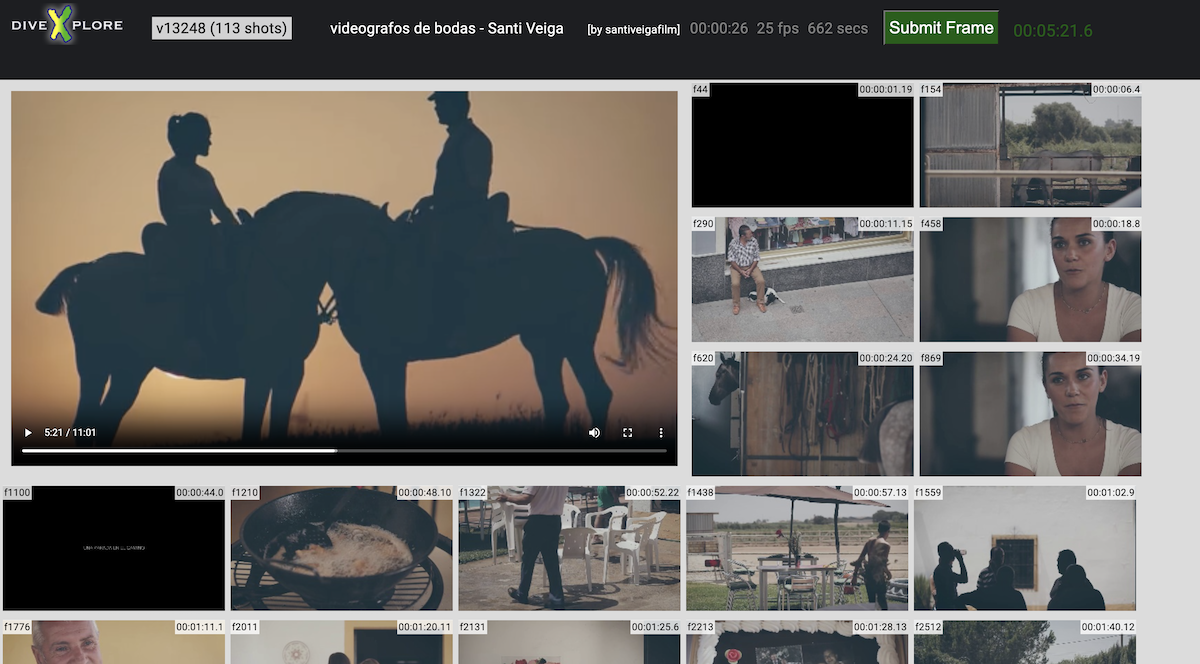}
  \caption{Video-view with player and shot list.}
  \label{fig:videoview}
\end{subfigure}
\hfill
\begin{subfigure}{0.9\textwidth}
  \includegraphics[width=\textwidth]{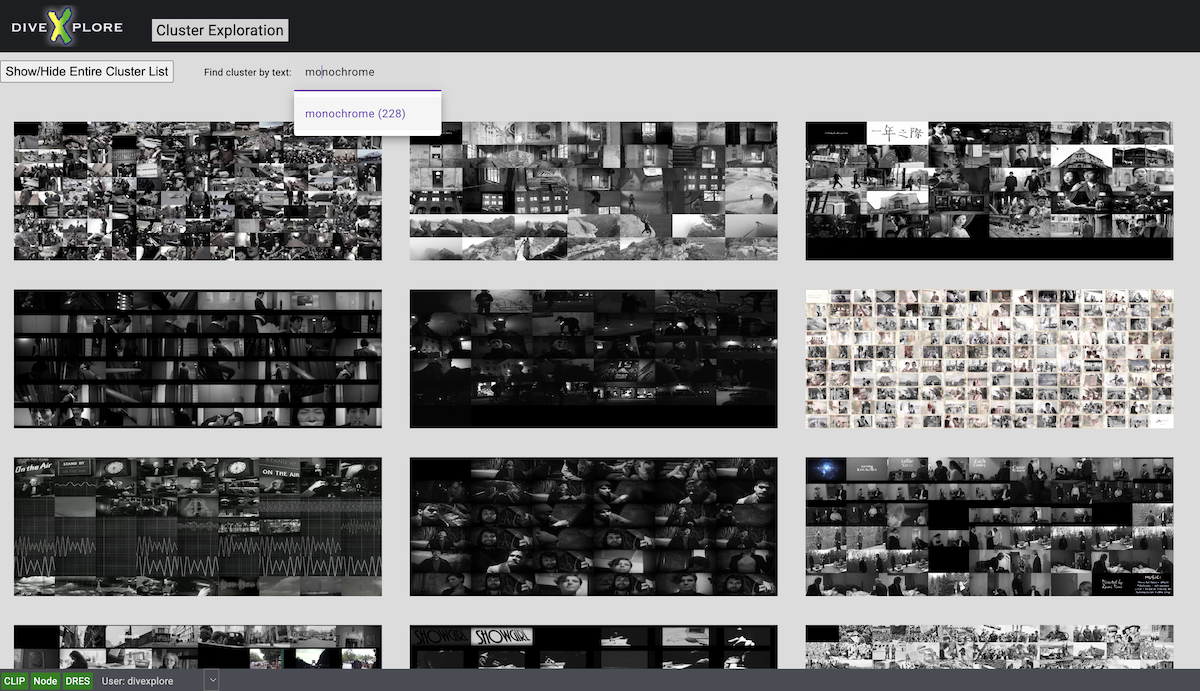}
  \caption{Exploration view of clusters with similar videos.}
  \label{fig:explorationview}
\end{subfigure}

\caption{Different views of diveXplore}
\end{figure}

The main user interface of diveXplore 2024 is shown in Figure~\ref{fig:userinterface}. It consists of a query selection (Free-Text, Temporal, Objects, etc.), a search bar, and a paging-based result list below. Whenever the user moves the mouse over a result, buttons for additional features  appear. For example, the user could inspect the corresponding video summary or the entire video with a video player, the shot list, and all meta-data (see Figure \ref{fig:videosummary}-\ref{fig:videoview}). The user could also perform a similarity search for the keyframe or send it directly to the VBS evaluation server (DRES) \cite{rossetto2021system-dres}. When the user moves the mouse horizontally over the top area of the result, different keyframes from the same video are presented as the mouse position moves; this should allow the user to very quickly inspect the video context of the keyframe. 
The result list is strongly optimized for keyboard-only use. The \textit{space} key can used to open and close the video summary for a keyframe, the \textit{arrow} keys are used to navigate between results and pages.
The search bar supports an expert mode that can combine several different search modalities (e.g., \textbf{c}oncepts, \textbf{o}bjects, \textbf{e}vents, \textbf{t}exts, etc.) with simple prefixes (e.g., \texttt{-c, -o, -e, -t} etc.). This way it is easy to combine search for embeddings with metadata filters, but also to enter temporal search by simply using less \texttt{<} and greater \texttt{>} characters.  
In case the user rather wants to explore, he/she can also use the Xplore button, which opens another view of large video clusters with similar content (see Figure \ref{fig:explorationview}), or a consecutive list of video summaries for all videos.

\section{Conclusion}

We introduce diveXplore for the VBS2024 competition. The system has been significantly improved in many ways, in particular the improved index of OpenCLIP embeddings, extracted with the model trained on LAION2B, and the flexible and efficient search possibilities with the distributed query server make our system very competitive. Last but not least, we would like to mention that we integrate specific content analysis for the surgical video dataset, which is crucial to find content in these videos with highly redundant content.

\section*{Acknowledgements}
This work was funded by the FWF Austrian Science Fund by grant P 32010-N38.

\newpage

% \begin{figure}
% \includegraphics[width=\textwidth]{fig1.eps}
% \caption{A figure caption is always placed below the illustration.
% Please note that short captions are centered, while long ones are
% justified by the macro package automatically.} \label{fig1}
% \end{figure}

\bibliographystyle{splncs04}
{
\scriptsize
\bibliography{refs.bib}
}

\end{document}